\documentclass[prb,preprint,amsmath,amssymb,superscriptaddress,floatfix]{revtex4}
\usepackage{epsfig}
\usepackage{graphicx}
\usepackage{amsmath,amssymb}
\usepackage{bm}
\usepackage[usenames, dvipsnames]{color}
\usepackage{multirow}
\usepackage{epstopdf}
\usepackage{ulem}

\def\ve{\varepsilon}

\def\unit#1{\ \mathrm{#1}}

\begin{document}

\title{Growth, strain and spin orbit torques in epitaxial NiMnSb films sputtered on GaAs}

\author{N.~Zhao}
\affiliation{London Centre for Nanotechnology, University College London, 17-19 Gordon Street, London, WC1H 0AH, United Kingdom}
\author{A.~Sud}
\affiliation{London Centre for Nanotechnology, University College London, 17-19 Gordon Street, London, WC1H 0AH, United Kingdom}
\author{H.~Sukegawa}
\affiliation{Research Center for Magnetic and Spintronic Materials, National Institute for Materials Science,
Tsukuba 305-0047, Japan}
\author{S.~Komori}
\affiliation{Department of Materials Science \& Metallurgy, University of Cambridge, 27 Charles
Babbage Road, Cambridge CB3 0FS, United Kingdom}
\author{K.~Rogdakis}
\affiliation{London Centre for Nanotechnology, University College London, 17-19 Gordon Street, London, WC1H 0AH, United Kingdom}
\author{K.~Yamanoi}
\affiliation{London Centre for Nanotechnology, University College London, 17-19 Gordon Street, London, WC1H 0AH, United Kingdom}
\author{J.~Patchett}
\affiliation{Cavendish Laboratory, University of Cambridge, Cambridge CB3 0HE, United Kingdom}
\author{J. W. A.~Robinson}
\affiliation{Department of Materials Science and Metallurgy, University of Cambridge, 27 Charles
Babbage Road, Cambridge CB3 0FS, United Kingdom}
\author{C.~Ciccarelli}
\email{cc538@cam.ac.uk}
\affiliation{Cavendish Laboratory, University of Cambridge, Cambridge CB3 0HE, United Kingdom}
\author{H.~Kurebayashi}
\email{h.kurebayashi@ucl.ac.uk}
\affiliation{London Centre for Nanotechnology, University College London, 17-19 Gordon Street, London, WC1H 0AH, United Kingdom}

\date{\today}

\maketitle
\textbf{We report current-induced spin torques in epitaxial NiMnSb films on a commercially available epi-ready GaAs substrate. The NiMnSb was grown by co-sputtering from three targets using optimised parameter. The films were processed into micro-scale bars to perform current-induced spin-torque measurements. Magnetic dynamics were excited by microwave currents and electric voltages along the bars were measured to analyse the symmetry of the current-induced torques. We found that the extracted symmetry of the spin torques matches those expected from spin-orbit interaction in a tetragonally distorted half-Heusler crystal. Both field-like and damping-like torques are observed in all the samples characterised, and the efficiency of the current-induced torques is comparable to that of ferromagnetic metal/heavy metal bilayers.}

\section{Introduction}

In crystals with broken inversion symmetry, a non-equilibrium spin polarisation can be generated by an electric current. This is termed the Edelstein effect\cite{Edelstein01,Levitov01} (or inverse spin-Galvanic effect\cite{Ivchenko01}), which was discovered in III-V compound semiconductors\cite{Silov01} that lack an inversion centre in their zinc-blende crystal unit cell. This current-induced non-equilibrium spin polarisation can be used to manipulate magnetic moments in non-centrosymmetric magnetic crystals\cite{Bernevig01,Manchon01,Chernyshov01,Endo01,Fang01,Kurebayashi01,Ciccarelli01,Wadley01} and also in low-symmetry interfaces\cite{Chen01,MacNeill01}. Akin to the Edelstein effect where an electric current generates a uniform spin polarisation across the entire sample, the spin-Hall effect (SHE)\cite{Dyakonov01,Hirsch01,Murakami01,Kato01,Wunderlich01,Sinova01} can generate spin currents from electric charge currents. This mechanism can be exploited to exert magnetic torques within multilayers of ferromagnetic and heavy high spin-orbit coupling metals\cite{Ando01,Liu02}. These two types of spin torques originate from the spin-orbit transport effects, often termed spin-orbit torques, have been widely studied in many different film stacks \cite{Miron00,Miron01,Kim01,Garello01,Emori01,Fukami01,Oh01,Fan01,Avic01,Ghosh01,Wu01,Skinner01,Cai01,Pospischil01,Cao01}. 

Although metallic multilayers are more common in the study of spin-orbit torques, non-centrosymmetric magnetic crystals do play a pivotal role in advancing the research field. Since spin-orbit torques in non-centrosymmetric magnetic crystals are generated within a single layer of a material and not at an interface, effect s due to interface s of the sample such as the SHE can be ruled out. In practice this makes the analysis of the experimental results simpler. For example, a sizable damping-like component of the spin-orbit torques in GaMnAs was discovered\cite{Kurebayashi01}, which might have been difficult to identify in metallic multilayers because of the SHE that gives a strong damping-like component in multilayers\cite{Ando01,Liu02}. In addition, the direction of the current-induced spin polarisation in non-centrosymmetric magnetic crystals is determined by the point group of the crystal\cite{Ciccarelli01,Hal01,Zelezny01}, which leads to a variety of torque symmetries that depend on the choice of material.    

In this paper, we characterise spin-orbit torques in sputter-deposited epitaxial NiMnSb thin-films. NiMnSb is a room temperature ferromagnet\cite{Felser01} and half-Heusler compound that have been studied for their predicted half-metallicity \cite{Otto01} as well as high magnetic tunability via stoichiometry\cite{Durrenfeld01}. Ciccarelli $et$ $al.$\cite{Ciccarelli01} discovered spin-orbit torques in epitaxial NiMnSb on lattice-matched InGaAs, which was tailor-made in a III-V molecular beam epitaxy (MBE) chamber.  In our study, we prepared epitaxial NiMnSb thin-films by sputtering on commercial epitaxial GaAs. Our structural analysis reveals a tetoragonal distortion in the epitaxial NiMnSb films due to the growth-induced strain from the GaAs substrates. By using current-induced ferromagnetic resonance (FMR) experiments, we identified spin-orbit torques arising from the strain-induced spin-orbit Hamiltonian. The NiMnSb thin-films exhibit five times larger torques than those measured in NiMnSb grown by MBE\cite{Ciccarelli01}, match the performance of ferromagnetic metal/heavy metal bilayers. Furthermore, epitaxial sputtered NiMnSb shows strong damping-like spin-orbit torques, which are absent in MBE NiMnSb \cite{Ciccarelli01}. 

\section{Sample growth and film characterisation}

A NiMnSb film was deposited on a GaAs(001) substrate by magnetron co-sputtering system with a base pressure of 1$\times$10$^{-7}$ Pa. Prior to each deposition, the GaAs substrate was cleaned with a 35\% hydrochloric acid (HCl) aqueous solution at room temperature for 9 minutes before being transferred into the sputtering chamber. Surface cleaning by annealling at 400$^{\circ}$C for 30 minutes was performed, followed by a sample cooling process down to the growth temperature of 250$^{\circ}$C. The NiMnSb films were grown by co-sputtering from Ni, Mn, and Sb targets where the sputtering powers for individual targets were tuned to control the sample stoichiometry close to the desired 1:1:1 ratio of each element. The NiMnSb deposition rate was set to 0.12 nm/s and the deposition was performed using an Ar partial pressure of 0.4 Pa. Mn and Sb were deposited by DC sputtering whereas Ni was deposited by RF sputtering. We grew samples with different NiMnSb thicknesses for growth optimization and for this particular study, we used a 17-nm-thick NiMnSb film to fabricate micron-scale bar devices. After the deposition, the films were post-annealed at 400$^{\circ}$C within the chamber. Finally, a 3-nm-thick MgO layer was deposited by RF-sputtering at room temperature. We confirm that the MgO layer acts as a capping layer by resistivity measurements of NiMnSb films. The composition of NiMnSb was measured as Ni$_{31}$Mn$_{33}$Sb$_{36}$ by inductively coupled plasma analysis of a 100-nm-thick reference film deposited on a thermally oxidized Si substrate at room temperature. The crystal structure was characterized by 4-axis X-ray diffraction (XRD) (Rigaku Smartlab) with Cu-K$\alpha$ radiation and a graphite diffracted beam monochromator. The magnetic hysteresis loops ($M$-$H$ loops) were measured using a vibrating sample magnetometer (VSM) at room temperature. The surface morphology was evaluated by atomic force microscopy (AFM). Figure 1 shows a summary of our film characterisation results. Figures 1 (a) and (b) trace the XRD profiles for out-of-plane (2$\theta-\omega$ scans) and in-plane rotations (2$\theta_{\chi}-\phi$ scans), respectively, with different incident directions.  Inset of Fig. 1 (b) shows the $\phi$ scan for a NiMnSb(222) plane. These traces confirm that the NiMnSb film is epitaxially grown on GaAs with cube-on-cube orientation relationship of GaAs(001)//NiMnSb(001) and GaAs[100]//NiMnSb[100], which is consistent with previously reported NiMnSb film on GaAs \cite{Gerhard01}. No secondary phases were detected in the profiles, indicating formation of a single C1$_{b}$ type NiMnSb phase. The 00l and 0ll peak angles (out-of-plane) are slightly lower than the l00 and ll0 angles (in-plane), respectively. Therefore, the NiMnSb film is tetragonally-distorted. Using the peaks' positions and the least-squares method, we obtained lattice constants with standard deviations $ a = b = 0.5937\pm0.0002$ nm and  $c = 0.5921\pm0.0002$ nm, giving $c/a=0.997\pm0.0002$. The sign of this distortion generated by growth induced strain is opposite to the lattice mismatch of their bulk lattice constants: the bulk value of the lattice constants are 0.593 nm for NiMnSb\cite{Fong01} and 0.5653 nm for GaAs\cite{Straumanis01}, for which we would expect the NiMnSb crystal grown on GaAs to have a compressive strain of about 4.9 \%. We speculate that misfit dislocations are periodically introduced near the GaAs/NiMnSb interface, which greatly influence the lattice relaxation. The residual tetragonal distortion that we measured in the X-ray scans can be determined by both residual sputter-deposition stress and lattice-mismatches. Figure 1 (c) displays the in-plane $M$-$H$ curves along [1$\overline{1}$0], [010], [110], and [100] in GaAs. The shape for the different crystallographic directions suggests the presence of both uniaxial and cubic magnetic anisotropies (we will discuss more detailed results on these below). The extracted saturation magnetisation is 668$\pm$17 mT, which is about 0.7 times smaller than the bulk value (930 mT)\cite{Ritchie01}. The saturation magnetisation of NiMnSb is sensitive to growth conditions \cite{Attema01,Orgassa01} and the stoichiometry of the NiMnSb. A similar reduction in epitaxial NiMnSb thin-films is reported by Kwon $et$ $al$\cite{Kwon01}. Possible causes for the reduction include: (i) a smaller concentration of Ni and Mn atoms with respect to Sb; (ii) the existence of a magnetically dead layer at the GaAs/NiMnSb and NiMnSb/MgO interfaces, and (iii) residual antisite defects in the C1$_b$ lattice. As shown in Fig. 1 (d), the AFM micrograph (1$\times$1 $\mu m^{2}$ area) of the NiMnSb surface reveals the mean average roughness of $Ra$ = 0.31 nm and the maximum peak-to-valley value of 4.8 nm from this characterisation. A key challenge of our growth optimisation was to maintain the high crystallinity of NiMnSb films, with minimal roughness. The crystallinity was improved by increasing the growth temperature up to 400$^{\circ}$C, at the expense of roughness. Therefore, we grew the NiMnSb at 250$^{\circ}$C with post annealing at 400$^{\circ}$C.

\section{Device fabrication and measurement techniques}

Each NiMnSb film has been cut into 5 mm square chips for device fabrication by optical lithography into bars. A device schematic together with the electric circuitry of the measurement setup is shown in Fig. 2(a). The NiMnSb bar width and length are 5 $\mu$m and 50 $\mu$m, respectively. We use the phenomenological Landau-Lifshitz-Gilbert (LLG) equation to describe the spin-orbit-induced magnetisation dynamics in our experiments:
\begin{equation}
        \frac{\partial\bm{M}}{\partial t} = -\gamma \bm{M} \times \bm{H}_\text{tot} + \frac{\alpha}{M_s} \left( \bm{M} \times \frac{\partial\bm{M}}{\partial t}\right)-\gamma \bm{M} \times \bm{h}_\text{so}.
            \label{eq:LLG}
    \end{equation}

Here $\gamma$ is the gyromagnetic ratio and $\alpha$ is the phenomenological Gilbert damping constant. $\bm{H}_\text{tot}$ is the total magnetic field vector and $\bm{h}_\text{so}$ is the effective spin-orbit magnetic field that drives the magnetic moments. Within a small precession angle approximation, where the magnetisation dynamics is in the linear excitation regime, we can write $\bm{M} = ($$M_\text{s}$$, m_{b} e^{i\omega t}, m_c e^{i\omega t})$, where $M_\text{s}$ represents the saturation magnetisation. 
We focus on the alternating in-plane angle (m$_b$(t)/M$_s$) since this motion leads to rectified voltages\cite{Kurebayashi01}. This magnetisation precession causes a time-varying resistance change originating in the anisotropic magnetoresistance (AMR): $R(\theta) = R_0 - \Delta R\cos^2(\theta)$, where $R_0$ and $\Delta R$ are magnetisation-angle independent and dependent terms respectively and $\theta$ is defined by the angle between the current flowing direction and static magnetisation orientation. This time-varying resistance change is mixed with a time-varying electric current component (which exert spin torques in our NiMnSb films), providing the experimentally measurable rectification voltages with the form of $V_{\text{dc}}=(I \Delta R m_b/2M_s) \sin 2\theta$ \cite{Kurebayashi01}. In this model, the real component of $V_{\text{dc}}$ is decomposed in the sum of symmetric and asymmetric Lorentzians\cite{Fang01}:
\begin{equation}
        \text{Re}\{V_\text{dc}\} = V_\text{sym}\frac{\Delta H^2}{(H - H_\text{res})^2 + \Delta H^2} + V_\text{asy}\frac{\Delta H(H -  H_\text{res})}{(H - H_\text{res})^2 + \Delta H^2}.
        \label{eq:realVdc}
    \end{equation}
Here, $H_\text{res}$, $H$ and $\Delta H$ are the resonance field for FMR, applied magnetic field and the linewidth of FMR peaks, respectively.
    \begin{eqnarray}
        V_\text{sym} (\theta) &=& \frac{I \Delta R \omega }{2\gamma\Delta H (2 H_\text{res} + H_1 + H_2)} \sin(2 \theta) h_z     \label{eq:Vsym} \\
        V_\text{asy} (\theta) &=& \frac{ I \Delta R (H_\text{res} + H_1)}{2\Delta H (2 H_\text{res} + H_1 + H_2)} \sin(2 \theta) (-h_x\sin\theta + h_y\cos\theta)     \label{eq:Vasy}
    \end{eqnarray}
are the weights that determine the lineshape of the resonance which will be used for quantifying the components of the spin-orbit effective field $h_x$, $h_y$ and $h_z$. $H_1$ and $H_2$ are:
    \begin{eqnarray}
        H_1 &=& M_s - H_{2\perp} + H_{2\parallel}\cos^2\left(\phi + \frac{\pi}{4} \right) + \frac{1}{4}H_{4\parallel}(3+\cos4\phi)   \\
        H_2 &=& H_{4\parallel}\cos 4\phi - H_{2\parallel}\sin 2\phi .
    \end{eqnarray}
$H_{2\perp}$, $H_{2\parallel}$ and $H_{4\parallel}$ represent the out-of-plane uniaxial, in-plane uniaxial and in-plane biaxial anisotropies respectively, and $\phi$ is the angle between the magnetisation vector $\mathbf{M}$ and the [100] crystallographic axis. 
In the experiments, we kept the frequency of the injected microwaves fixed whist sweeping the external dc magnetic field. We also perform similar experiments by adding ac magnetic modulation, as described in the Supplementary Material, which provides results consistent with the dc magnetic field experiments. Figure 2 (b) displays typical spin-orbit FMR data together with best fit curves obtained using Eq. (\ref{eq:realVdc}). The fit curves show a good quality of fitting, confirming the validity of the macrospin rectification model discussed above for our measurements.

\section{FMR analysis and spin-orbit torques}

By using the equations above, we are able to quantify the magnetic and spin-orbit parameters through a series of FMR measurements. Figure 3 (a) shows $V_{\text{dc}}$ measured for a bar oriented along the [110] direction as a function of $H$ and $\phi$. We can clearly identify a resonance in each scan, where $H_\text{res}$ depends on $\phi$ as shown in Fig. 3 (b) and is used to deduce magnetic anisotropy parameters using our model. As summarised in Table I, we can identify that both uniaxial and bi-axial (crystalline) anisotropies are present in the epitaxial NiMnSb films, consistent with VSM results (Fig. 1(c)). These magnetic anisotropies are comparable with earlier reports\cite{Koveshnikov01, Gerhard01}. Gerhard $et$ $al.$ quantified epitaxial NiMnSb films grown on a InP/(In,Ga)As substrate with different Mn concentrations and observed a large range of anisotropy changes in both terms\cite{Gerhard01}, in particular showing that the uniaxial magnetic anisotropy becomes lager when the in-plane lattice constant deviates from the stoichiometic value. Our samples have growth-induced strain that modifies the lattice constant away from the bulk value. Although it is difficult to make direct quantitative comparison between NiMnSb on different substrates, we can note that the magnitude of $H_{2\parallel}$ and $H_{4\parallel}$ observed in our study matches fairly well with that measured by Gerhard $et$ $al.$ It is also possible that when epitaxial ferromagnetic films with growth-induced strain are lithographically patterned, uniaxial strain relaxation takes place, leading to a patterned-induced uniaxial magnetic anisotropy\cite{Wenisch01,King01}. We would expect a similar anisotropy term in our NiMnSb film where the growth-induced compressive strain ($c/a$ = 3.5$\times 10^{-4}$) might relax when patterning a bar. However, when we compare bars patterned along the [110] and [1$\overline{1}$0] directions we do not identify a clear presence of the patterned-induced term in the anisotropy within our detection limit since both devices exhibit very similar $H_{2\parallel}$ and $H_{4\parallel}$ values. It is worthwhile to note that a uniaxial anisotropy term is still present in epitaxial NiMnSb films even though the lattice constant of the two main axes (i.e. $a$ and $b$) match. This suggests the presence of additional symmetry lowering in the film plane from four-fold into two-fold, which cannot be captured by our X-ray probe on crystallograpic parameters. A similar result was previously observed in GaMnAs epitaxial thin-films where it was attributed to a uniaxial anisotropy which can be understood by proposals of uniaxial distribution of Mn atoms that breaks the symmetry between the [110] and [1$\overline{1}$0] orientations\cite{Kopecky01, Birowska01}. This can potentially lower an effective point group down to $2mm$, allowing this uniaxial anisotropy to exist\cite{Birowska01}.The Gilbert damping coefficient of our NiMnSb film is characterised as 0.007 using broad-band FMR experiments.
\begin{center}
\begin{table}[h!]
\caption[S1]{The magnetic parameters deduced from dc and ac FMR measurements. }
\begin{tabular}{|c|c|c|c|c|c|c|c|c|}
    \hline
  Sample \#& \multicolumn{2}{ |c| }{1} & \multicolumn{2}{ |c| }{2} & \multicolumn{2}{ |c| }{3} & \multicolumn{2}{ |c| }{4} \\\hline
  Direction & \multicolumn{2}{ |c| }{[1$\overline{1}$0]} & \multicolumn{2}{ |c| }{[110]} & \multicolumn{2}{ |c| }{[1$\overline{1}$0]} & \multicolumn{2}{ |c| }{[110]}  \\\hline
  Modulation & dc & ac & dc & ac & dc & ac & dc & ac \\\hline
  $H_{2\parallel}$ [mT] & -4 & -5 & -8 & -7 & -5 & -6 & -10 & -9\\\hline
  $H_{4\parallel}$ [mT] & 5 & 5 & 6 & 5 & 7 & 4 & 7 & 6\\\hline
  $M_\text{eff}$ [mT] & 646 & 645 & 669 & 645 & 681 & 652 & 664 & 628\\\hline
\end{tabular}
\end{table}
\end{center}
The angular dependence of $V_\text{sym}$ and $V_\text{asy}$ reveals the spin-orbit torque parameters\cite{Fang01,Kurebayashi01}. The field-like components of spin-orbit torques are associated to dissipative mechanisms\cite{Bernevig01,Manchon01} and are depend on the non-equilibrium distribution function of conduction electrons with specific spin textures in the Fermi surface. For NiMnSb, these components are parametrised by the in-plane components of the spin-orbit field, $h_x$ and $h_y$, which produce an asymmetric lineshape of the resonance in $V_\text{dc}$. By fitting the angular dependence of $V_\text{asy}$, we extract the magnitudes of $h_x$ and $h_y$ for each sample. Figures 4 (a) and (b) include the angular dependence of $V_\text{asy}$ (black dots) for both [1$\overline{1}$0] and [110] samples respectively, together with a best fit obtained by using Eq. (\ref{eq:Vasy}). Both angular dependences show a clear sin2$\theta$cos$\theta$ symmetry, indicating that $h_y$ is the dominant component of the spin-orbit field. Furthermore, the sign flip between the [110] and [1$\overline{1}$0] samples, indicates that $h_y$ has the opposite signs for a current flowing along the [110] and [1$\overline{1}$0] directions. These observations are consistent with the previous work on MBE NiMnSb films\cite{Ciccarelli01}. By using the symmetry of the spin textures in momentum space for each spin-orbit term, we are able to decouple the Dresselhaus and Rashba components of the effective fields; the sign of the spin states for a momentum along the [110] and [1$\overline{1}$0] directions is opposite to each other in the Dresselhaus spin texture, while it is in the same direction in the Rashba spin texture as depicted in Figs. 4 (c). Hence, the effective fields from the Dresselhaus and Rashba terms ($h_D$ and $h_R$) can be calculated as 
$h_D = (h_{y,110}- h_{y,1\overline{1}0})/2$ and $h_R = (h_{y,110}+ h_{y,1\overline{1}0})/2$ respectively, where $h_{y,110}$($h_{y,1\overline{1}0}$) is the $h_y$ component extracted for the bar oriented along the [110] ([1$\overline{1}$0]) direction. In our experiment it is found that the Dresselhaus component is much greater than the Rashba one. When a zinc-blende crystal is distorted by elongating one of [100], [010] and [001] orientations, the crystal point group is reduced from $\overline{4}$3m into $\overline{4}$2m. In a system with $\overline{4}$2m symmetry, the following spin-orbit term ($H_D$) is allowed, giving rise to the 2D Dresseuhaus spin texture as shown in Fig.4 (c) (gray arrow):
\begin{equation}
H_D = \beta[J_xk_x(\epsilon_{yy}-\epsilon_{zz} )+ c.p.].
  \label{eq:Hd}
   \end{equation}
Here, $\beta$, $J_i$, $k_i$ and $\epsilon_{ii}$ are the coefficient for this term, $i$ component of angular momentum of conduction carriers, the wavevector along the $i$ direction and a diagonal element of the strain tensor in NiMnSb respectively; $c.p.$ represents cyclic permutation. When a zinc-blende crystal is under shear strain, meaning that an off-diagonal component of the strain tensor is non-zero, the point group is reduced to mm2, which allows the following Rashba-type spin orbit term and the spin texture is illustrated in Fig.4 (c) (purple arrow):
\begin{equation}
H_R = \alpha_R[(J_x k_y-J_y k_x)\epsilon_{xy}+ c.p.].
\label{eq:Hr}
\end{equation}
Here, $\alpha_R$ and $\epsilon_{ij}$ represents the coefficient for this Rashba term and an off-diagonal element of the strain tensor, respectively. Experimentally deduced values of $h_D$ and $h_R$ should be proportional to the strength of $H_D$ and $H_R$ respectively \cite{Fang01,Kurebayashi01}. The growth-induced strain identified by our X-ray characterisation produces finite values of $(\epsilon_{yy}-\epsilon_{zz})$ as well as $(\epsilon_{zz}-\epsilon_{xx})$, hence generating $h_D$. However, we could not identify the presence of the off-diagonal strain, suggesting that this strain component is minute. Therefore, it is reasonable that the dominant effective field is $h_D$ over $h_R$. 

We notice that in our samples there is a sizable component of $V_\text{sym}$. In our macrospin model, $V_\text{sym}$ leads to the presence of damping-like torques which can also be interpreted as being induced by an out-of-plane component of the spin-orbit field $h_z$. This is because the magnetisation precession induced by $h_z$ is phase-shifted by $\pi/2$ with respect to the one generated by the aforementioned field-like torques, leading to a lineshape change into Lorentzian symmetric $V_\text{sym}$. In GaMnAs it was demonstrated that the current-induced damping-like torques originate from the Berry curvature of the electronic band structures in GaMnAs\cite{Kurebayashi01}. Contrary to the previous report\cite{Ciccarelli01}, in our case we observe a non-negligible $V_\text{sym}$ components for both [110] and [1$\overline{1}$0] samples as shown in Figs. 4 (a) and (b). In order to analyse the damping-like torque component, we allow $h_z$ in Eq. (\ref{eq:Vsym}) to be angular dependent as $h_z (\theta) = h_0 + h_1 \text{sin}\theta + h_2 \text{cos}\theta$. By comparing the general forms of the field-like and damping-like torques, $i.e.$ $\bf{m \times}(m \times s) $ and $\bf{m \times}h $, we can interpret that the z component of $\bf{h}$ can be written as: $h_z = m_y s_x - m_x s_y$. Using the definition of $\theta$ in our analysis, we find that $(m_x,m_y)=M_\text{s}(\text{cos}\theta,\text{sin}\theta)$ are the in-plane components of the magnetisation, while ($ s_x $, $ s_y $) are the components of the current-induced spin-polarisation. Substituting this into the previous equation, we find $h_z = M_\text{s}(s_x \text{sin}\theta - s_y \text{cos}\theta)$. Therefore, for a $s_x$ ($s_y$) spin polaristaion, $h_z (\theta)$ has sine (cosine) symmetry. Clearly, our observation of $V_\text{sym}$ in Figs. 4 (a) and (b) displays $\text{sin}2\theta$ $\text{cos}\theta$ symmetry, suggesting that $h_z (\theta)$ must have $\text{cos}\theta$ symmetry upon the in-plane rotation of $M$. This, together with the aforementioned analysis, leads to the conclusion that the predominant contribution of the damping-like torques is from $s_y$ spin polarisation in momentum space. In Fig. 4 (e) we plotted $h_0$, $h_1$ and $h_2$. A sign change in $h_2$ between the [110] and [1$\overline{1}$0] samples is consistent with a sign flip of $s_y$ therefore with the Dresselhaus symmetry of the spin texture in NiMnSb, as also found in our previous analysis of the field-like torques. Finally, we show the linear dependence of the rectified longitudinal voltage $ V_\text{dc} $ on microwave power in Fig. 4(f). This dependence is consistent with our spin-texture model, since $V_\text{sym} \propto I^2 \propto P$ where $h_i (i=x,y,$ and $z) \propto I$ in Eqs. (3) and (4). 
\begin{center}
	\begin{table}[b!]
	\caption{SOT effective fields for various material systems. All of them have been scaled by the current density of $j = 1 \times 10^{10} $ A/$\text{m}^2$. The following unit is used for effective fields: $h_{FL,DL}/j$ $[\text{mT}/(10^{10} \text{A}/\text{m}^2)]$. NiMnSb measured in this paper are marked with *. Note that a more completed list can be available in Ref. \cite{Manchon02}. }
		\begin{tabular}{|c|c|c|c|}
			
			\hline    
			Structure type &Name&	 $h_{FL\|}$/j &	 $h_{DL\|}$/j \\ \hline
			Bulk Ferromagnets &	(Ga,Mn)As \cite{Kurebayashi01}	&-2.01&	-1.27\\ \hline
			&	NiMnSb\cite{Ciccarelli01}	&-0.06&	\\ \hline
			&	NiMnSb*&	-0.31&	-0.48	\\ \hline
			Nonmagnetic metals &	Pt/Co/AlO$_{x}$ \cite{Garello01}&	0.4&	-0.69\\ \hline
			&	Ti/CoFe/Pt \cite{Fan01} &-0.03&	 0.32	\\ \hline
			&	Ta/CoFeB/MgO \cite{Avic01}	&-0.21&	0.32	\\ \hline
			&	Pd/Co/AlO$_{x}$ \cite{Ghosh01}&	0.07	&0.13 \\ \hline
			Antiferromagnets     &	IrMn$_{3}$/CoFeB/MgO \cite{Wu01} & 0.07	& -0.18 \\ \hline
			Semiconductors     &	(Ga,Mn)As/Fe \cite{Skinner01} & 0.03	& -0.03 \\ \hline
			Topological insulators     &		Mn$_{0.4}$Ga$_{0.6}$/Bi$_{0.9}$Sb$_{0.1}$ \cite{Khang01} & 	& -230   \\ \hline
		\end{tabular}
		
	\end{table}
\end{center}

To complete our analysis, it is useful to compare the values of the SOT in our sputtered NiMnSb films to those reported in previous works. We use an effective current-induced field size ($h_{FL}$ and $h_{DL}$ for field-like and damping-like fields respectively) per normalised current density of $j = 1 \times 10^{10} $ A/$\text{m}^2$ as figure of merit for this analysis. Our NiMnSb films show $h_{FL}=0.31$ mT and $h_{DL}=0.48$ mT. These are larger than $h_{FL}$ in MBE-grown NiMnSb $h_{FL}=0.06$ mT\cite{Ciccarelli01}. This result has a technological interest since sputtering techniques are compatible to industrial processes and less costly. Furthermore, we also highlight that our $h_{FL/DL}$ values are comparable to those reported in heavy-metal/ferromagnet bilayer systems as listed in Table II. For completeness, we also list the spin-orbit field size measured in sputtering-grown topological insulators, which is as large as 230 mT at $j = 1 \times 10^{10} $ A/$\text{m}^2$ as reported by Khang et al.\cite{Khang01} very recently.

\section{Conclusions}
In this study, we present measurements of spin-orbit torques in sputter-deposited epitaxial NiMnSb on a GaAs substrate. X-ray characterisation reveals that NiMnSb grows epitaxially with a cube-on-cube orientation relationship of GaAs(001)//NiMnSb(001) and GaAs[100]//NiMnSb[100]. A growth-induced strain is present, which distorts the lattice constants as $ a = b = 0.5937\pm0.0002$ nm and  $c = 0.5921\pm0.0002$ nm, giving $c/a=0.997\pm0.0002$. We characterise the spin-orbit torques by two current-induced FMR methods and find that the torques originate from the spin-orbit interaction within the tetragonally distorted NiMnSb crystal. Furthermore, both field-like and damping-like torques are found to be present in our samples, which is different from spin-orbit torques in MBE-grown NiMnSb. We find that the effective field generation efficiency of our sputtering-grown NiMnSb is comparable to those reported from heavy-metal/ferromagnet bilayer systems.

\section*{Acknowledgments}

We are grateful to Seiji Mitani for his technical supports for sputtered film preparation, as well as Kota Hanzawa and Hidenori Hiramatsu for their efforts to grow NiMnSb films.

\newpage
\begin{figure}[h]
\begin{center}
\includegraphics[scale=0.7]{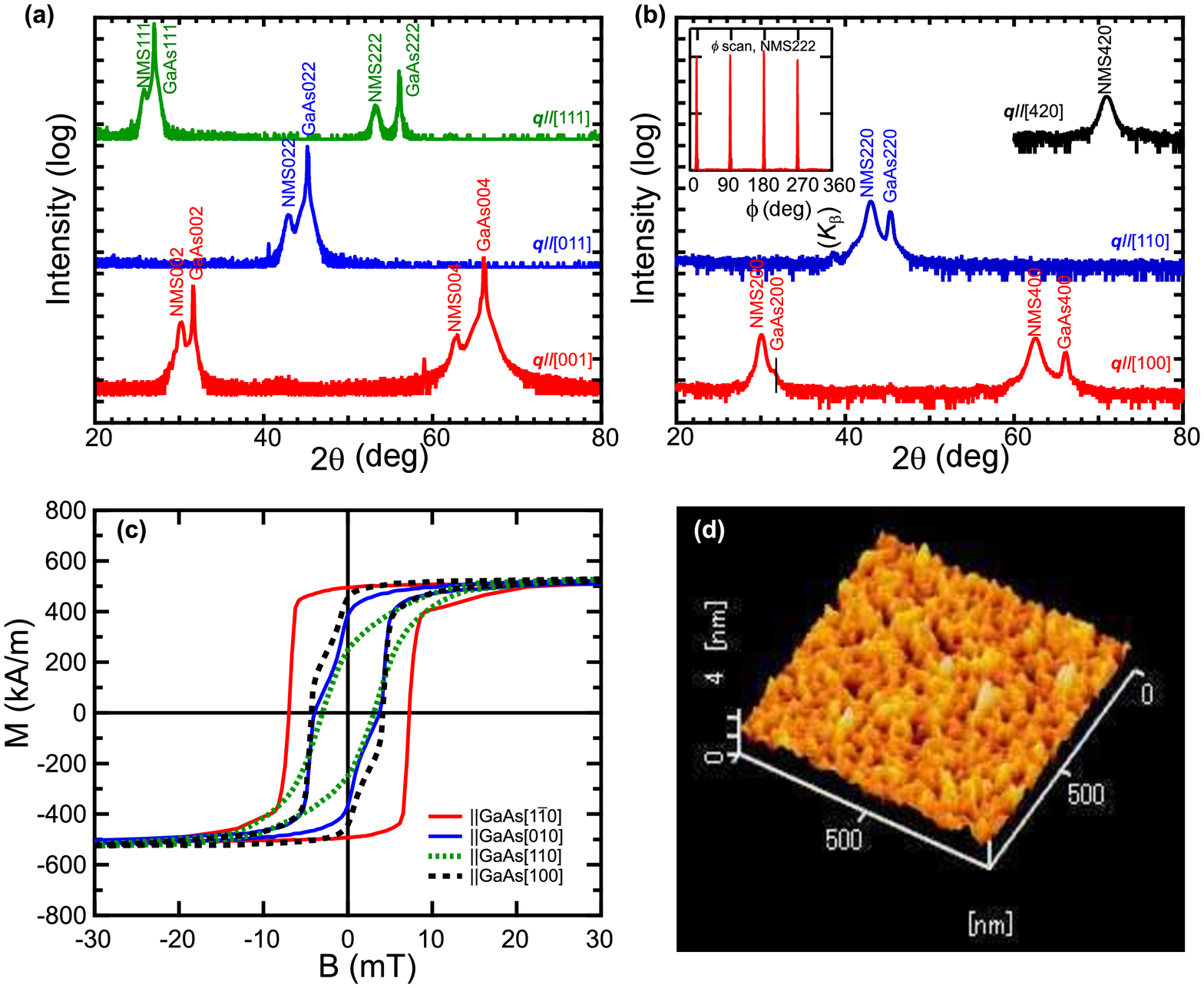}
\end{center}
\caption{(a),(b) XRD profiles for GaAs(001) sub./NiMnSb (NMS) 17 nm/MgO 3 nm film. (a) Out-of-plane and (b) in-plane profiles along various scattering vectors $q$ of GaAs. Inset of (b) shows NMS222 $\phi$ scan profile (pole figure). (c) In-plane $M$-$H$ curves along GaAs[1$\overline{1}$0], [010], [110], and [100]. (d) An AFM image of 17 nm of the NiMnSb film used in the present study.}
\end{figure}

\newpage
\begin{figure}[h]
\begin{center}
\includegraphics[scale=0.4]{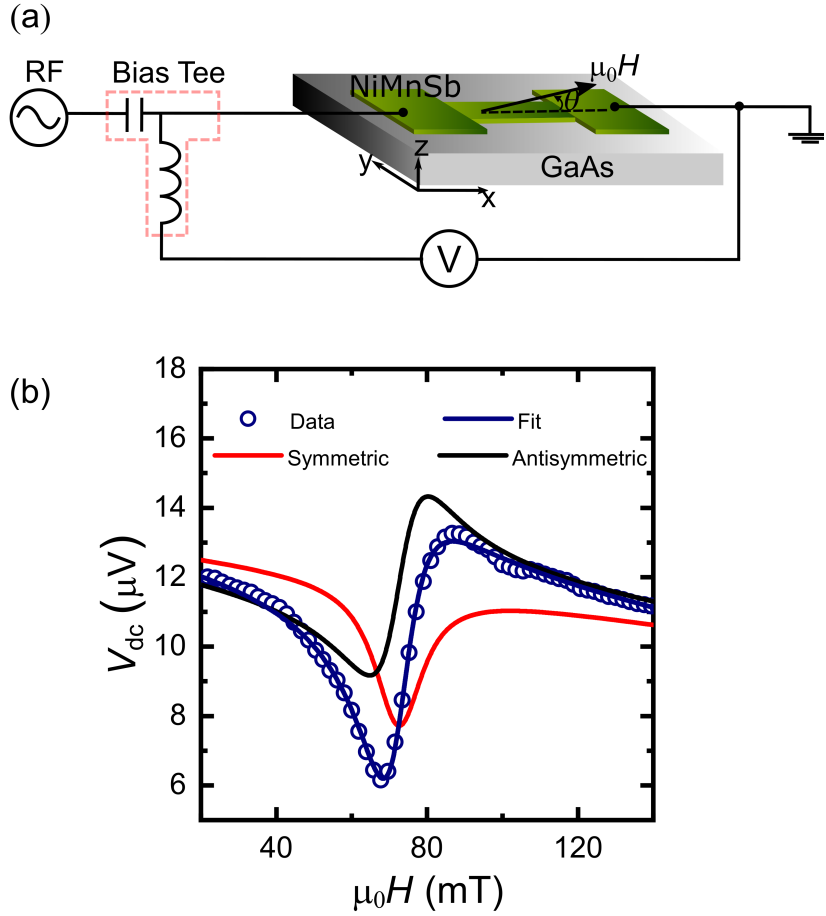}
\end{center}
\caption{(a) Schematic of the measurement set-up using this study with circuit configuration. (b) Typical FMR voltages measured at microwave excitation frequency of 7 GHz with the external magnetic field applied along $\theta$ = 60$^{\circ}$ in the [110] device. Dots are experimental data points and curves are produced by best fit parameters using Eq. (\ref{eq:realVdc}) and voltage offsets that are constant with field and linear to field.}
\end{figure}

\newpage
\begin{figure}[h]
\begin{center}
\includegraphics[scale=0.6]{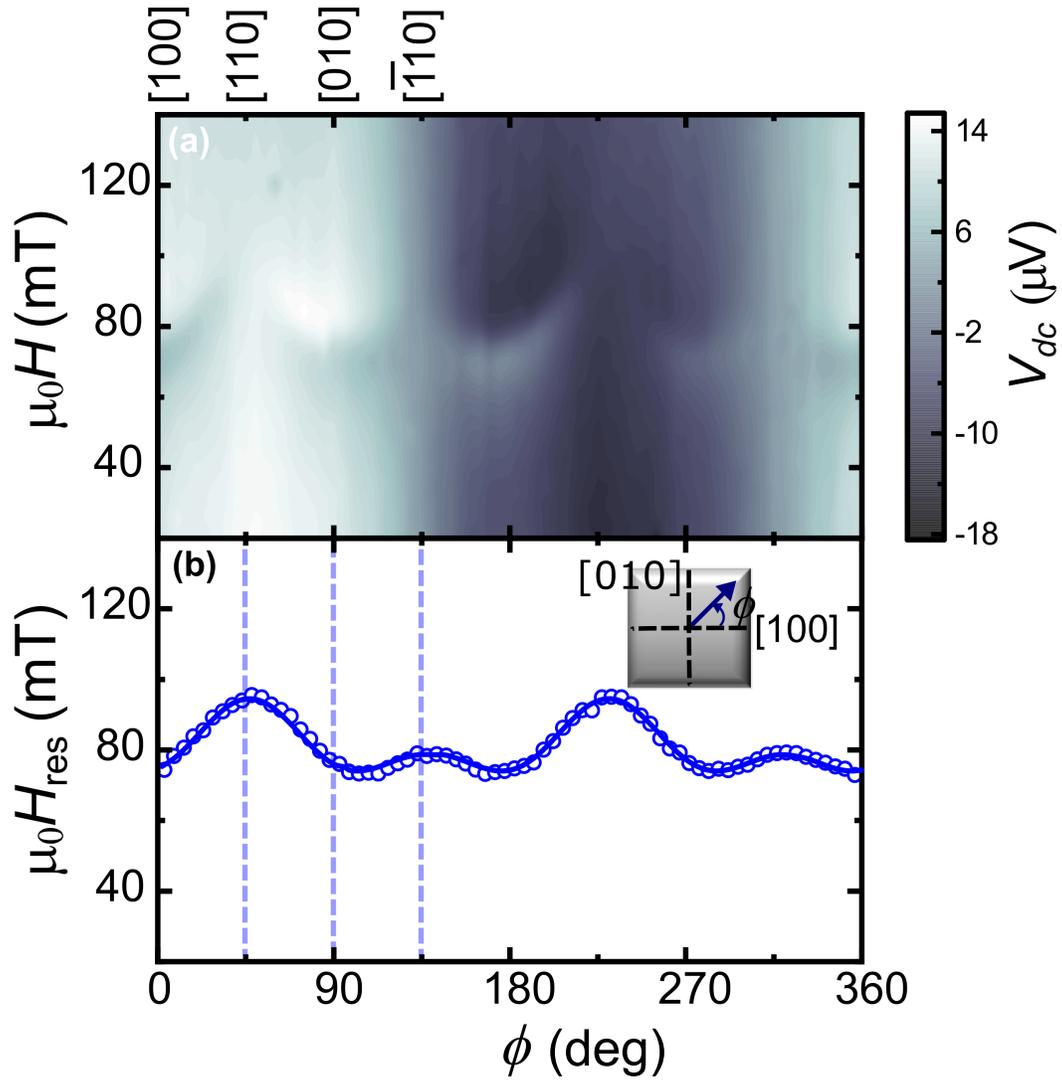}
\end{center}
\caption{(a) Two-dimensional color plot of FMR voltages  $V_\text{dc}$ measured from in-plane rotational scan of external magnetic field as a function of applied magnetic field and angle. We use the microwave excitation frequency of 7 GHz. (b) FMR field $H_\text{res}$ as a function of the in-plane crystallograhic angle. We plot results from fitting FMR scans.}
\end{figure}

\begin{figure}[h]
\begin{center}
\includegraphics[scale=0.30]{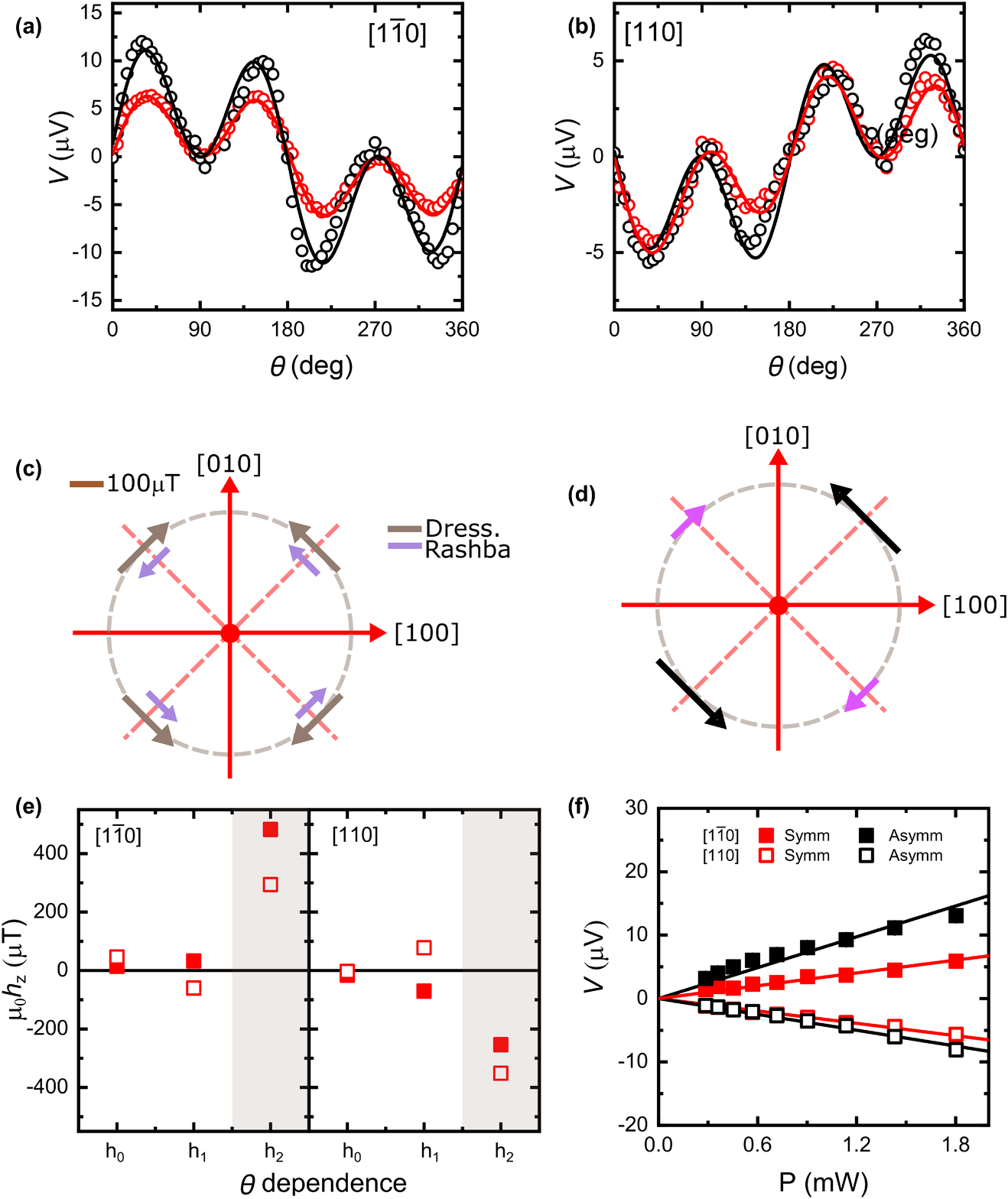}
\end{center}
\caption{ (a-b) Angle dependence of the symmetric (empty red circles) and anti- symmetric (empty black circles) components of the resonance measured for bars patterned along the (a) [1$\overline{1}$0] and  (b) [110] crystal directions. The current density in the bar is $ 10^{10}$A/$\text{m}^2 $ and measured at microwave excitation frequency of 7 GHz. (c-d) Plot of the magnitude and direction of the effective fields for (c) Dresselhaus and  Rashba  type of symmetry and  (d) the sum of the two components. (e) $\theta $ dependence of effective $h_z$ field. (f) Plot of magnitude of voltage for symmetric and antisymmetric components as a function of injected microwave powers.}
\end{figure}
\pagebreak
\widetext
\begin{center}
\textbf{\large Supplementary Material for "Growth, strain and spin orbit torques in epitaxial NiMnSb films sputtered on GaAs"}

\end{center}
\renewcommand{\theequation}{S\arabic{equation}}
\renewcommand{\thefigure}{S\arabic{figure}}
\setcounter{figure}{0}

\makeatletter
\renewcommand{\theequation}{S\arabic{equation}}
\renewcommand{\thefigure}{S\arabic{figure}}
\renewcommand{\bibnumfmt}[1]{[S#1]}
\renewcommand{\citenumfont}[1]{S#1}

\def\ve{\varepsilon}
\def\unit#1{\ \mathrm{#1}}
\font\myfont=cmr12 at 16pt
\section{Sample characterisation}

\subsection{Anisotropic magnetoresistance}
The response of charge carriers to the relative orientation change between magnetisation $\mathbf{M}$ and electrical current $\mathbf{I}$ in magnetic materials is called anisotropic magnetoresistance (AMR) that can be phenomenologically defined as:
\begin{equation}
R_{AMR}\equiv (R(\theta )-\bar{R})/\bar{R}
\end{equation}
where $R(\theta )$ is the longitudinal resistance with respect to the angle $\theta$ between $\mathbf{M}$ and $\mathbf{I}$, $\bar{R}$ is the resistance average over $\theta$. The measurements of AMR are presented in this section as well as a discussion into its sign and magnitude. In our approach to investigate AMR, an in-plane saturating magnetic field is employed to rotate the magnetisation in NiMnSb along bar directions of [1$\overline{1}$0], [110], [100] and [010]. The angular dependence of longitudinal AMR along four different bar directions are shown in Fig.S1 for the 17nm NiMnSb films with 5$\mu$m bar width. We noticed a $cos(2\theta )$ dependence of AMR, by re-writing Eq.(S1) into $R_{AMR}=Ccos(2\theta )$, the AMR coefficient $C$ along [1$\overline{1}$0] and [110] bar direction is extracted with a value of 0.07$\%$ which is comparable to the AMR coefficient measured by Ciccarelli et al in NiMnSb\cite{Ciccarelli02}. This can be explained by two AMR components, one depending on the angle between the electric current and magnetisation directions, and the other depending on the angle between the electric current and crystallographic orientations \cite{Ciccarelli02}. Here we add that GaMnAs also shows similar two components in AMR \cite{Rushforth01,Rushforth02}.
    \begin{figure}[htb]
	\begin{center}
		\includegraphics[width=0.6\textwidth]{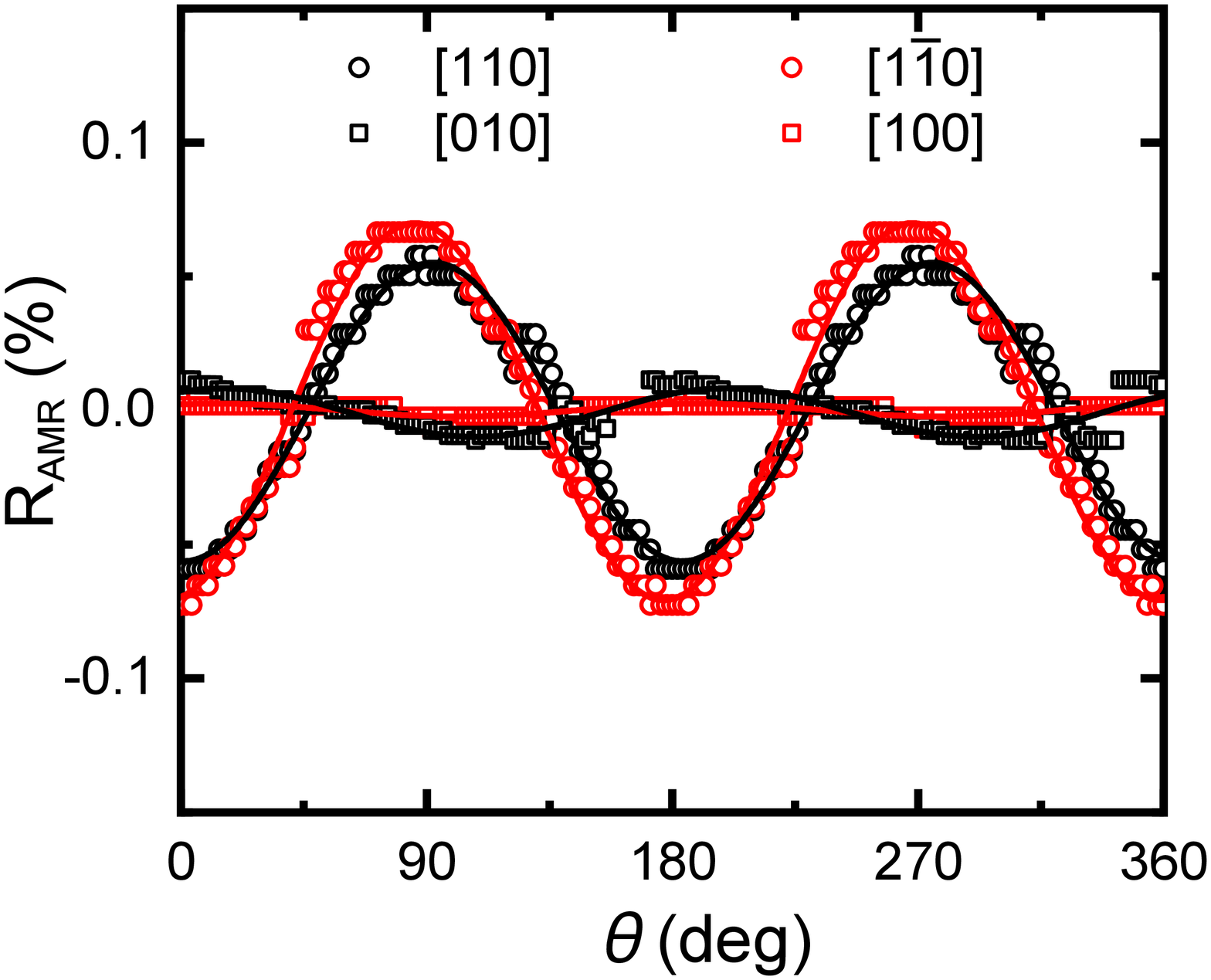}
	\end{center}
	\caption{AMR measured in [1$\overline{1}$0], [110], [100] and [010] oriented bars.}
\end{figure}
For a system exhibiting both uniaxial and cubic anisotropies, the longitudinal AMR along four bar directions can be described as\cite{Ciccarelli02}:
\begin{equation}
\frac{\Delta \rho _{L,100/010(\theta )}}{\overline{\rho }}=C_{C}cos4\theta+(C_{I}+C_{IC})cos2\theta\pm C_{U}sin2\theta
\end{equation}
\begin{equation}
\frac{\Delta \rho _{L,1\bar{1}0/110(\theta )}}{\overline{\rho }}=-C_{C}cos4\theta+(C_{I}-C_{IC})cos2\theta\pm C_{U}cos2\theta 
\end{equation}
where $\rho _{L}$ is longitudinal resistivity, $\bar{\rho}$ is the average longitudinal resistivity. The four terms $C_{I}$, $C_{IC}$, $C_{C}$ and $C_{U}$ represent noncrystalline, mixed crystalline, cubic crystalline and uniaxial crystalline AMR parameters.

These parameters are extracted by fitting the AMR results measured along [100] and [010] bar directions using Eq.(S2) and AMR measured along [1$\overline{1}$0] and [110] by Eq.(S3), we also define the AMR ratio as $\chi \equiv \frac{\rho _{L(m\parallel j)}-\rho _{L(m\perp j)}}{\bar{\rho}}$ to compare the sign and magnitude of AMR among different materials. The AMR parameters from our measurements and previous NiMnSb study\cite{Ciccarelli02} are summarised in Table I below. A negative AMR ratio $\chi$ is observed in both NiMnSb samples which is consistent with the prediction of negative $\chi$($\rho _{L(m\parallel j)}<\rho _{L(m\perp j)}$) in half-metallic materials \cite{Kokado01}. The magnitude is comparable to experimentally determined $\chi$ at around -0.10\% in NiMnSb under room temperature \cite{Kwon02}. The sign of AMR ratio is the same as in (Ga,Mn)As\cite{Vyborny01} while most transition metals have positive $\chi$\cite{Kokado01}($\rho _{L(m\parallel j)}>\rho _{L(m\perp j)}$). The AMR is dominated by $C_{I}$ (noncrystalline term) and $ C_{IC}$ (mixed crystalline tern) which explains the leading $cos(2\theta )$ term of AMR shown in Fig.S1. However the $C_{I}$ and $ C_{IC}$ term are not completely cancelled out which results in an observable but at least ten times smaller AMR in [100] and [010] bar directions than that of [1$\overline{1}$0] and [110] directions. These small AMR values significantly affect our spin-orbit torque (SOT) ferromagnetic resonance (FMR) experiments, since the size of the rectification voltages is proportional to the AMR value. As a result, it is not possible to observe sizable FMR peaks across the entire magnetic field angle for [100] and [010] bar directions.
\begin{center}
	\begin{table}[htb]
				\caption{Fitted AMR parameters using phenomenological AMR equation.}

		\begin{tabular}{|c|c|c|c|c|c|c|}
						\hline    
			Sample&	$\chi _{100}($\%$) $&$\chi _{110}($\%$) $$ $&$C_{I}$&$C_{IC}$&$C_{C}$&$C_{U}$\\ \hline
			NiMnSb&	-0.01	&-0.15&	-0.032&	0.031& 0.003& 0.009\\ \hline
			NiMnSb\cite{Ciccarelli02}&-0.01	&-0.12&	-0.028	&0.033& 0.001& -0.005\\ \hline
		\end{tabular}
\end{table}
\end{center}

\subsection{Microwave calibration}
The microwave power used for current-induced SOT-FMR experiments was up to 28 dBm from the microwave source, however there is a large proportion of power reflection from the device due to a significant impedance mismatch between the microwave line and the devices (a typical device resistance is a few k$\Omega$). In order to quantify the current-induced effective fields $h_{\mathrm{eff}}$, we employed a bolometric microwave power calibration techniques where a sample resistance change due to Joule heating is used to quantify the microwave power injected into the samples \cite{Fang02,Kurebayashi02,Phu02}. To compare the change in resistance due to Joule-heating by direct current (dc) or microwave power, first a dc sweep is performed with no microwave power while the resistance is being measured as shown in Fig. S02(a). The resistance change is compared to the same measurements but using microwave currents as a heat source as shown in Fig. S2(b). By using two resistance change data sets, we are able to calibrate the microwave currents in the sample.

\begin{figure}[htb]
	\begin{center}
		\includegraphics[width=0.48\textwidth]{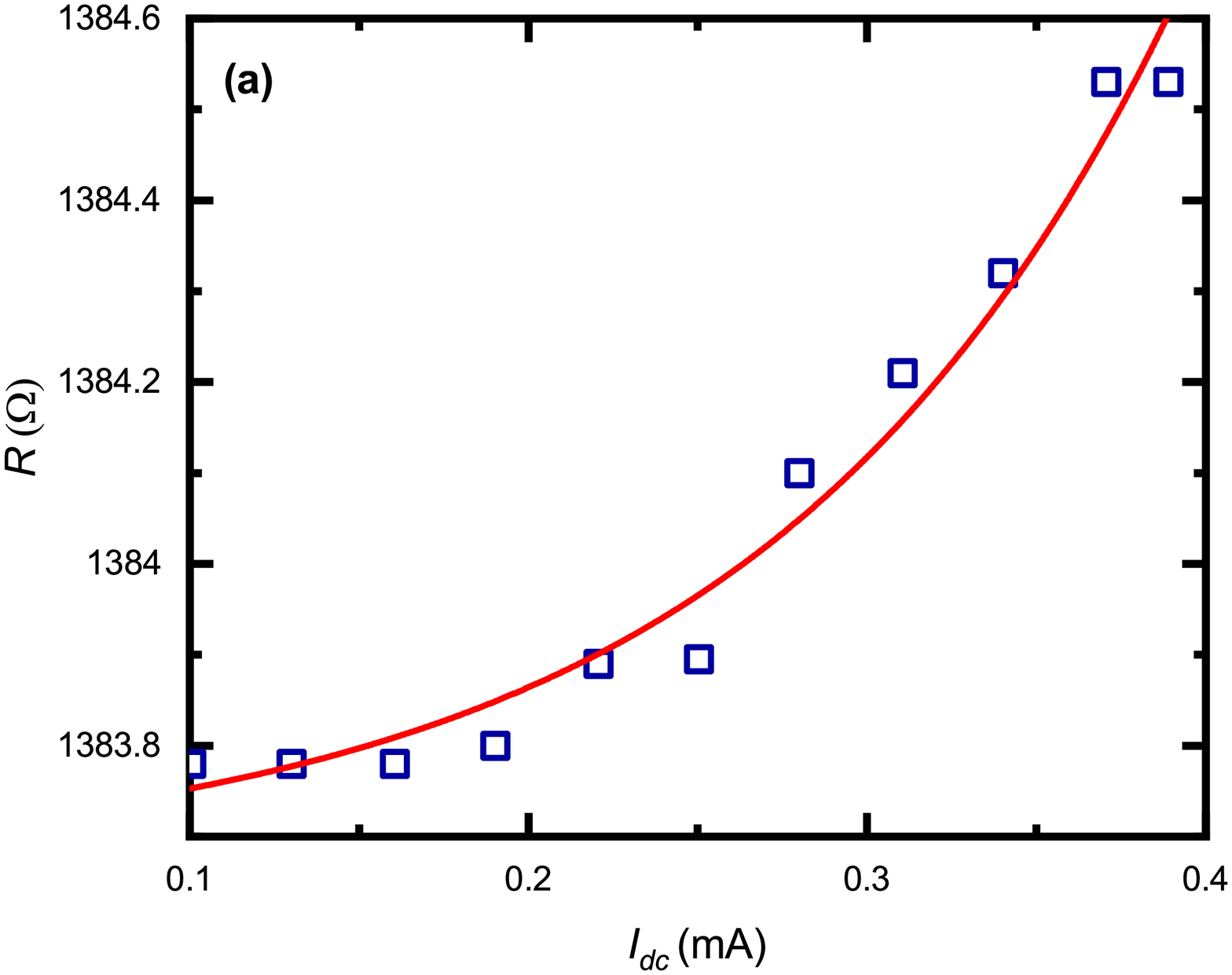}
		\includegraphics[width=0.48\textwidth]{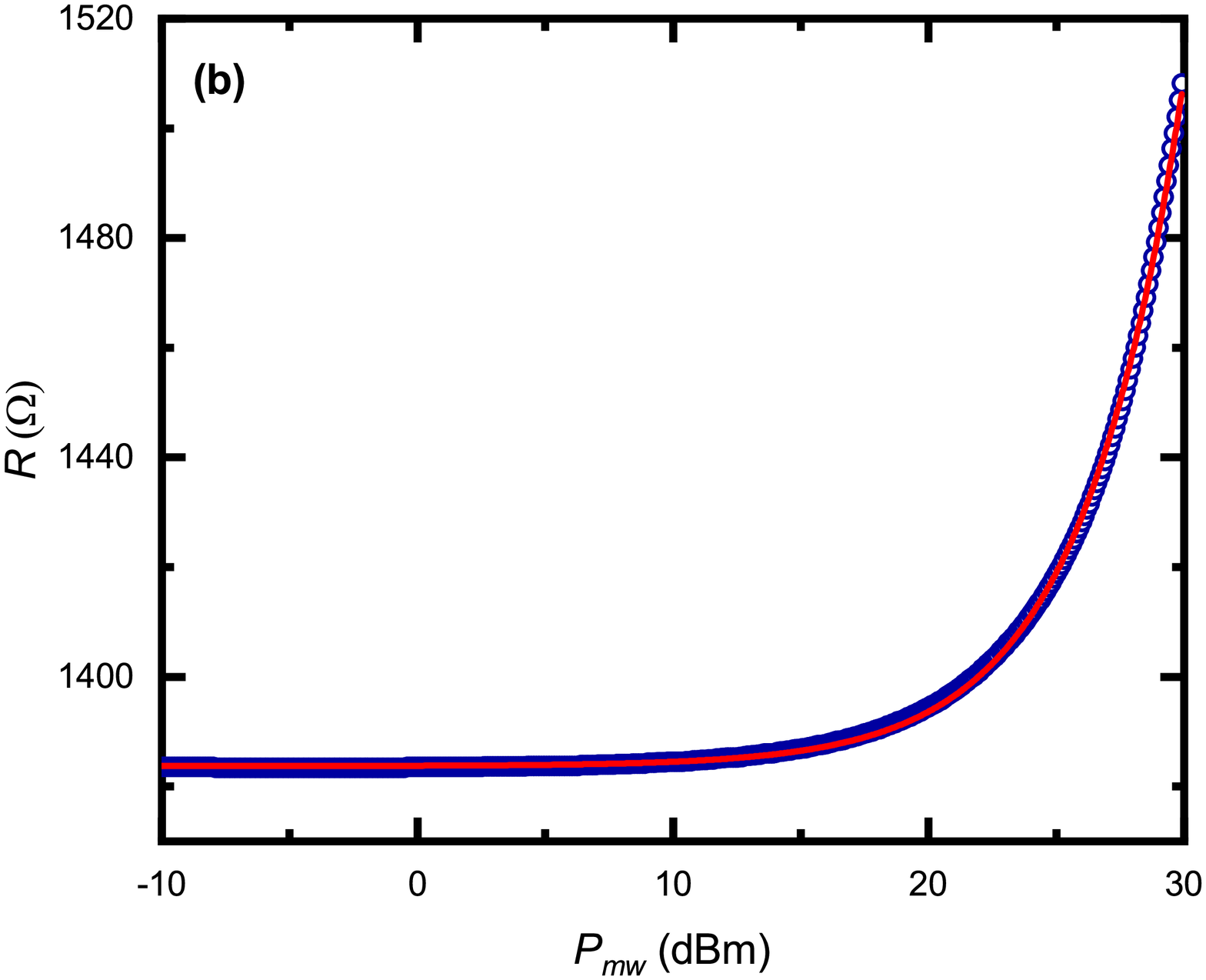}
	\end{center}
	\caption{Heating calibration for a bar oriented along [110] direction. (a) dependence of resistance on direct current without microwave power. (b) dependence of resistance on microwave power without dc current.}
\end{figure}

\section{FMR analysis and spin-orbit torques by ac magnetic field modulation}

In order to support our measurements in pulsed-modulated microwave currents, we here show our results to characterise the same material parameters using continuous microwave currents with alternating magnetic field components. This can be achieved by a pair of Helmholtz coils to provide an alternating magnetic field upon a very large static magnetic field. Due to this field modulation, the expected rectification voltage $V_{\mathrm{ac}}$ has the following form:
\begin{equation}V_{a c}=\frac{d V_{d c}}{d H} h_{a c}\end{equation}
where ${h}_{\text {ac}}$ is the magnitude of ac magnetic field modulation and by substituting Eq.(2) from the main manuscript, we obtain:
\begin{equation}V_{a c}=-V_{s y m} h_{a c} \frac{2\left(H-H_{r e s}\right) \Delta H^{2}}{\left.\left(\left(H-H_{r e s}\right)^{2}+\Delta H^{2}\right)\right)^{2}}-V_{a s y} h_{a c} \frac{\Delta H\left(\left(H-H_{r e s}\right)^{2}-\Delta H^{2}\right)}{\left(\left(H-H_{r e s}\right)^{2}+\Delta H^{2}\right)^{2}}\end{equation}

The first term gives now the anti-symmetric lineshape and the second term produces the distorted symmetric lineshape. We used this equation to fit FMR curves from our field modulation experiments. Fig. S3(a) displays typical FMR lineshape measured by field modulation and its best fit curves using Eq. (S5). The in-plane FMR voltage measurements are performed to investigate angle dependence of $V_{\mathrm{ac}}$ as well as to extract magnetic parameters. We plot a 2D color plot for $V_{\mathrm{ac}}$ as a function of magnetic field and angle $\phi$ in Fig. S3(b). We confirm that $ {H}_{\mathrm{res}}$ extracted this ac measurement method is consistent with our previous dc measurements as shown in Fig. S4. Furthermore, we analysed the angular dependence of $V_{ac}$ in the same way as $V_{\mathrm{dc}}$ presented as Fig. 4 in the main text. We observed clear sin2$\theta$cos$\theta$ symmetry for this measurement method and the sign reversal for devices with [1$\overline{1}$0] and [110] current directions. For full comparison, we have created Table II which lists all the magnetic and spin-orbit parameters extracted using the ac techniques. These extracted values agree with those extracted from dc measurement techniques, strongly suggesting that the waveform of microwaves (pulse-modulated or continuous) does not cause any artifact to alter the parameters.

\begin{figure}[htb]
	\begin{center}
		\includegraphics[width=0.8\textwidth]{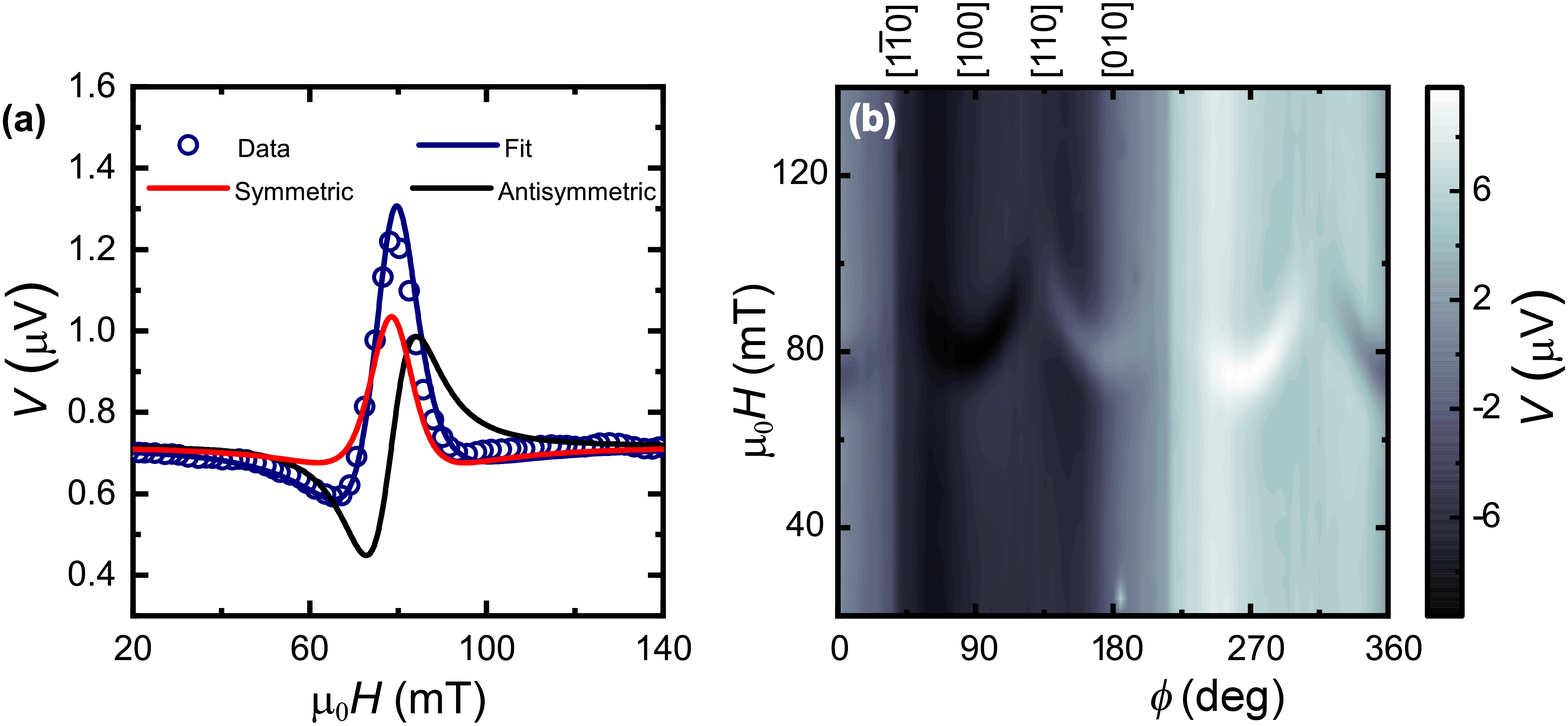}
			\end{center}
	\caption{(a) typical FMR curves measured by ac modulation. (b) two-dimensional colour plot of FMR voltages  $V_{\mathrm{ac}}$ as a function of applied magnetic field.}
\end{figure}

\begin{figure}[htb]
	\begin{center}
		\includegraphics[width=0.5\textwidth]{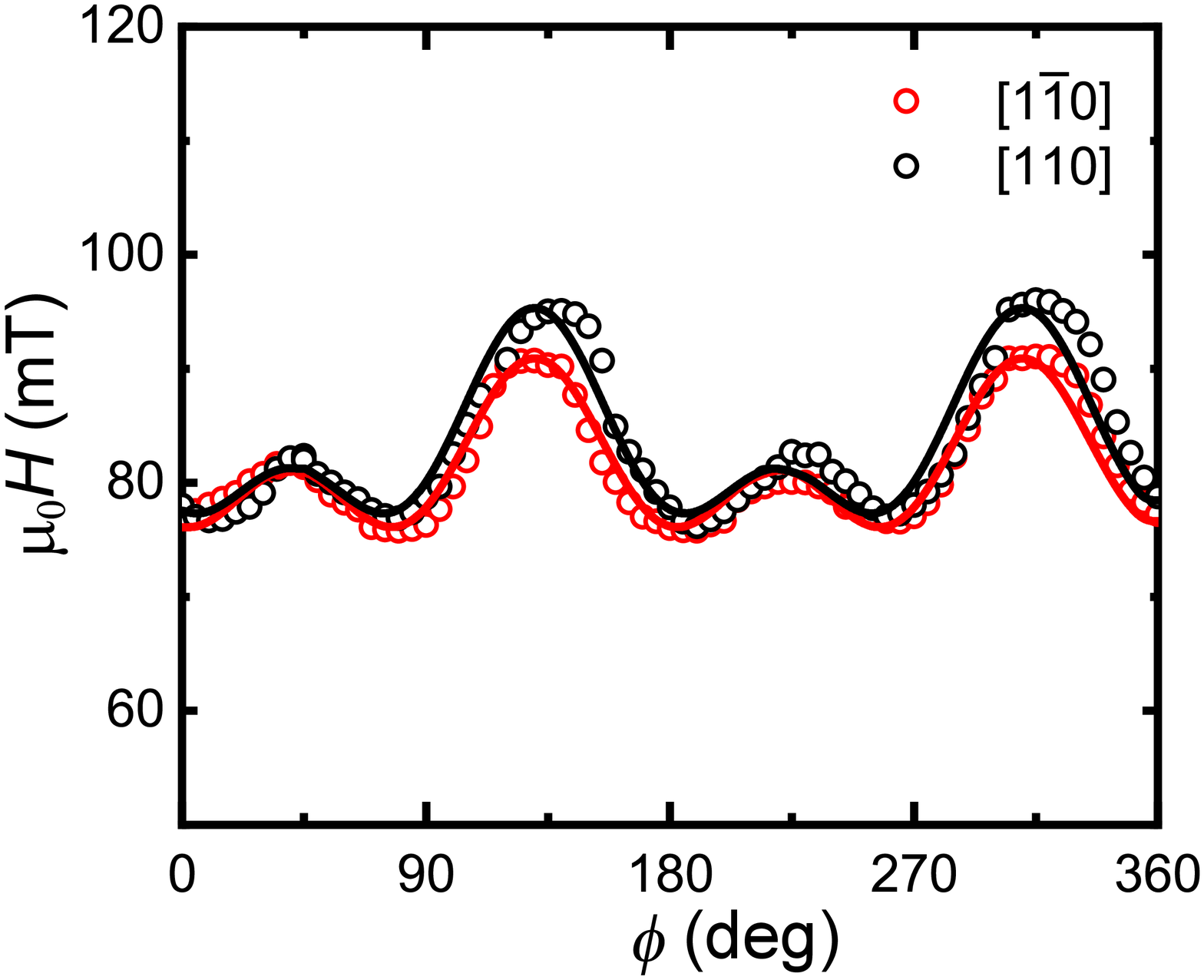}
	\end{center}
	\caption{FMR field $H_{\mathrm{res}}$ as a function of the in-plane crystallographic angle.}
\end{figure}

\begin{center}
	\begin{table}[htb]
	\caption{Magnetic parameters and effective fields deduced from FMR measurements. }
		\begin{tabular}{|c|c|c|c|c|}
			
			\hline    
			Sample \#&	1&	2&	3&	4\\ \hline
			Direction&	[1-10]	&[110]&	[1-10]&	[110]\\ \hline
			$H_{2\|}$   [mT]&	-5	&-7&	-6	&-9\\ \hline
			$H_{4\|}$  [mT]&	5&	5&	4	&6\\ \hline
			$M_{eff}$ [mT]&	645&	645&	652&	628\\ \hline
			$h_x$ [$\mu$T]&	-18	&-25&	-13	&-15\\ \hline
			$h_y$ [$\mu$T]&	316	&-151&	215	&-286\\ \hline
			$h_0$ [$\mu$T]&	13&	-30	&33	&20\\ \hline
			$h_1$ [$\mu$T]&	7&	-38	&-56&	98\\ \hline
			$h_2$ [$\mu$T]	&438&	-222&	360&	-295\\ \hline
		\end{tabular}
	\end{table}
\end{center}

\begin{figure}[htb]
	\begin{center}
		\includegraphics[width=0.8\textwidth]{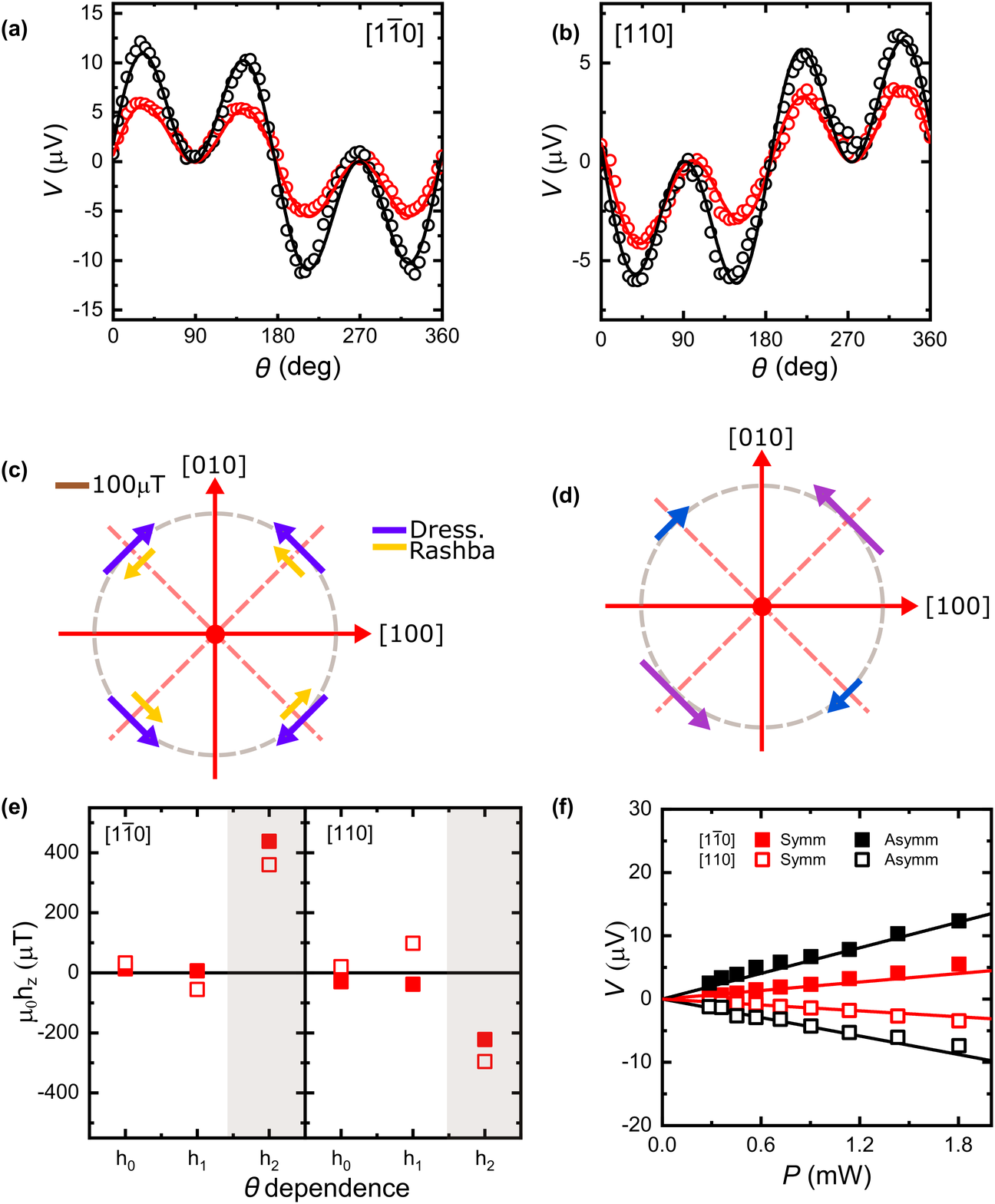}
			
	\end{center}
	\caption{(a-b) angular dependent of symmetric and anti-symmetric components $V_{ac}$ measured along [110] and [110] bar
		directions
		(c) the sum of Dresselhaus and Rashba type of symmetry.
		(f) angular dependence of effective ${h}_{z}$ field.}
\end{figure}

\end{document}